\documentclass[useAMS,usenatbib,rotating]{mn2e}

\usepackage{amssymb}     \usepackage{amsmath}    \usepackage{amsfonts}
\usepackage{xspace}    \usepackage{longtable}    \usepackage{rotating}
\usepackage{lscape}

\newcommand{\HI}{\mbox{\sc      H{i}}}     
\newcommand{\sex}{\mbox{\sc SExtractor}}

\newcommand{\msun}{\mbox{$M_\odot$}}    
              
\newcommand{\kms}{\mbox{km    s$^{-1}$}}

\newcommand{\miriad}{{\sc Miriad}\xspace}

\title{NOIRCAT   --  The   Northern   HIPASS  Optical/IR   Catalogue}
\author[O. I.  Wong et al.]   {O. I. Wong,$^{1,2,3}$ R.  L. Webster,$^1$
V.  A. Kilborn,$^{4}$  M. Waugh,$^{1}$  and  L. Staveley-Smith$^{5}$\\
$^1$School of Physics, University  of Melbourne, VIC 3010, Australia\\
$^2$Australia Telescope  National Facility, CSIRO, PO  Box 76, Epping,
NSW 1710, Australia\\ $^3$ Department of Astronomy, Yale University, New Haven, CT 06520-8101, USA\\$^4$  Centre for Astrophysics \& Supercomputing,
Swinburne University of Technology,  P.O. Box 218, Hawthorn, VIC 3122,
Australia\\ $^5$  School of Physics, University  of Western Australia,
35 Stirling Hwy, Crawley, WA 6009, Australia\\ }

\begin{document}

\date{Accepted ***. Received ***; in original form ***}

\pagerange{\pageref{firstpage}--\pageref{lastpage}} \pubyear{2007}

\maketitle
 
\label{firstpage}

\begin{abstract}
We present  the Northern HIPASS  Optical/InfraRed CATalogue (NOIRCAT),
an optical/near-infrared counterpart  to the Northern HIPASS catalogue
(NHICAT).  Of  the 1002 sources  in NHICAT, 655 ($66\%$)  have optical
counterparts with matching optical  velocities.  A further 85 (8\%)
  sources have  optical counterparts  with matching
velocities from previous radio emission-line surveys.  We  find a
correlation  between  the  gas  and  stellar content  of  the  NOIRCAT
sources.  Our \HI-selected sample of isolated galaxies also present a wider 
range in near-infrared (NIR) colours than previous optically-selected studies
 of regular, isolated galaxies. All \HI\ detections in optically unobscured
fields could be matched with either a NED optical counterpart, or a galaxy
visible in POSSII or DSS images.  However, as over 200 of these matched
galaxies have no velocity information, further follow-up observations
are needed to confirm the matches, and hence confirm or deny the existence 
of dark galaxies in this dataset.

\end{abstract}


\section{Introduction}
Primordial gas  clouds are postulated  to exist in the  Local Universe
due to the slow collapse of material about small density perturbations
present in the matter density field (after the epoch of recombination)
which  have not  reached the  threshold density  needed to  form stars
\citep{giov89}.  $N$-body simulations  of galaxy formation using   the     
Cold Dark Matter cosmological model \citep{moore99,klypin99}
predicted a significant number of small dark matter halos.   Since dark 
matter halos exist around most galaxies, small dark matter haloes
are assumed to exist around dwarf galaxies. However, current 
observations find the number of dwarf galaxies to be significantly less
than the predicted number of small dark matter halos.  Without   
modifying the large-scale properties of these  models, it may be possible  
for small dark matter halos to exist and galaxies not to have been 
observed if star 
formation had been suppressed  in these dark  halos.  Two  
possibilities exist:  (i) the halos  may contain  gas but  star  formation 
is suppressed; (ii)  the halos do not contain gas.   Current reionisation
models of the Universe predict the latter as they find that 
the gas from 95\%  of the low-mass systems  ($M_{\rm{virial}} \leq 10^8$ 
\msun\ or $v_{\rm{circ}} \leq 20$ \kms) appears to have been 
photoevaporated during the epoch of reionisation \citep{susa04}.

\citet{schneider96} suggested  that \HI\ surveys can be  used to probe
the regions  of the Local Universe  where stars have  not formed since
most  surveys conducted  with  optical telescopes  are biased  against
objects such as low surface brightness (LSB) galaxies and the proposed
dark galaxies.   There are also numerous  objects such as  NGC 2915, a
small  blue  compact dwarf  galaxy  in  optical  wavelengths, with  an
enormous  envelope   of  \HI\   extending  beyond  5   Holmberg  radii
\citep{meurer96}.  Hence, blind all-sky \HI\ surveys (i.e.\ HIPASS) may
reveal  a  large  undiscovered  population of  gas-rich  LSB  galaxies
\citep{disney76} as well as other gas-rich dwarf galaxies.

The \HI\  Parkes All-Sky Survey  (HIPASS) is the  largest blind
\HI\ survey, covering  $71\%$ of the total sky  using the Parkes Radio
Telescope\footnote{The  Parkes  telescope  is  part of  the  Australia
Telescope  which  is  funded  by  the Commonwealth  of  Australia  for
operation as a National  Facility managed by CSIRO.}.  Northern HIPASS
surveys    the    entire   sky    within    the   declination    range
$+2^{\circ}<\delta<+25.5^{\circ}$,  whereas Southern HIPASS  covers the
entire Southern sky south of a declination of $+2^{\circ}$.  The Northern
HIPASS  catalogue \citep[NHICAT;][]{wong06}  and  the Southern  HIPASS
catalogue  \citep[HICAT;][]{meyer04} detected  1002 and  4315 galaxies
respectively, based solely on the  \HI\ content.  Here, we present the
{\bf{N}}orthern     HIPASS    {\bf{o}}ptical     and    near-{\bf{IR}}
{\bf{cat}}alogue   (NOIRCAT)---a   catalogue  of   optical   and
near-infrared counterparts to the \HI\ galaxies in NHICAT.  NOIRCAT is
analogous to  the HIPASS Optical  Catalogue \citep[HOPCAT;][]{doyle05}
which is a catalogue of optical counterparts for HICAT.

There are many  theoretical arguments for \citep{davies06,verde02} and
against \citep{taylor05} the existence  of dark galaxies which will be
further discussed in Section 3.   To avoid confusion, we define a dark 
galaxy to be
an optically dark, isolated \HI\ source, with no neighbouring galaxies
and  no stars.   Previous discoveries of  ``dark'' galaxies were
either high  velocity clouds \citep{kilborn00} or gas clouds associated
with  optical galaxies \citep{schneider83,ryder01}.  One of the best
dark galaxy candidate is the isolated \HI\ cloud, HI1225+02
\citep{giov89}.  Higher spatial resolution mapping of HI1225+01 
showed two dynamically-distinct components \citep{chengalur95}; one
of which is associated with a LSB dwarf, while the other (south-western
cloud) is starless down to at least 27.2 magnitudes in the $I$-band
\citep{turner97}.


The most  recent unconfirmed dark galaxy  candidate is GEMS\_N3783\_2,
an isolated  region of  \HI\ gas with  no visible  optical counterpart
located   within  the  NGC   3783  galaxy   group  \citep{kilborn06}.
\citet{kilborn06} concluded that  GEMS\_N3783\_2 was formed during the
interaction  of NGC  3706 and  ESO 378-G003.   However,  the projected
separation  of GEMS\_N3783\_2  and ESO  378-G003  is 450  kpc with  no
obvious   \HI\  bridge   or  tail   structures.   Hence   deeper  follow-up 
observations  are  needed to  confirm and uncover  any further  \HI\
 or optically-faint counterparts remaining in this system.

Previous  blind  HI surveys,  including  HIPASS,  have  not found  any
evidence for  the existence of  dark galaxies.  Even  though the main purpose
of NOIRCAT is  to provide  complementary optical/near-infrared  data to 
NHICAT, NOIRCAT will also be able to provide an independent search for
dark galaxies.

Section 2 describes the construction  of  NOIRCAT  and  the
 properties  of the  catalogued galaxies.   Discussion 
of the scientific implications can be found in Section 3 and 
Section 4 provides a summary of our results.

\section{NOIRCAT}

This  section describes  the method  used to  produce NOIRCAT  and the
 properties  of NOIRCAT.   To  probe the  existence of  dark
 galaxies,   sources   with  no   optical   velocity  matches   within
 $7.5\arcmin$ of the HIPASS centre will be further analysed in Section
 2.3.
\begin{table*}
\caption{Definition of flags in the processing of NOIRCAT and the total
number of Northern HIPASS sources within each flag category.}
\label{flagtab}
\footnotesize{
\begin{center}
\begin{tabular}{clc}
\hline 
Flag  & Definition & Total \\  
\hline 
1 &Single optical  velocity match with  2MASS counterpart & 414 \\  
2& Single  optical velocity  match without 2MASS counterpart & 126 \\  
3 & Multiple optical velocity  matches where all matches  also have  2MASS counterparts & 63\\  
4 &Multiple  optical velocity matches where one  or more matches are without  2MASS counterparts & 52\\ 
5a & No  optical velocity  match  but with  higher angular  resolution \HI\  velocity match & 85\\  
5b &No  velocity match but positional matches available to NED galaxies & 221 \\
5c & No velocity or positional matches to any galaxies listed in NED & 41\\ 
\hline
\end{tabular}
\end{center}}
\end{table*}
\begin{figure}
\begin{center}
\includegraphics[scale=0.4,angle=270]{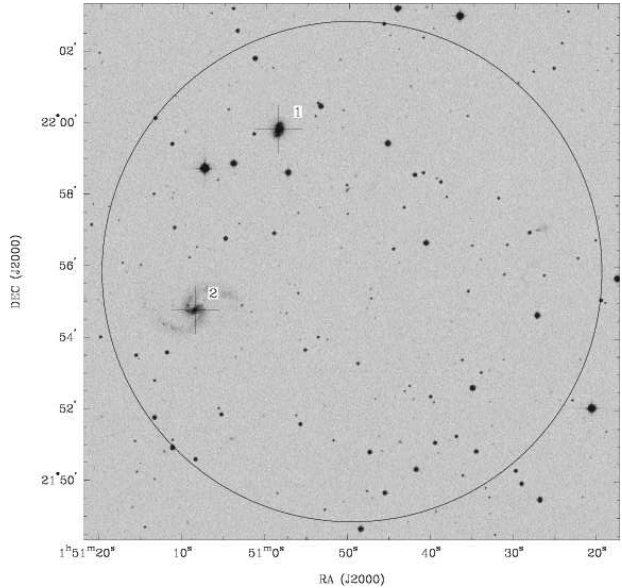}
\caption{Screenshot of an example field window used during the interactive inspection.  The field centre is of HIPASSJ0150+21, a Flag 3 detection. The circle shows the 7.5\arcmin\ radius from the HIPASS centre which is in the centre of the field.  All matches found in NED are marked by a `+' and a number corresponding to an optical source listed in the text window.}
\label{sample}
\end{center}
\end{figure}

\subsection{The construction of NOIRCAT}

There  have been  several  methods with  which  catalogues of  optical
counterparts of  HIPASS   samples have   been   produced.   Both
\citet{kilborn02} and \citet{ryan02} used NED\footnote{The NASA/IPAC  
Extragalactic Database (NED) is operated by the Jet Propulsion Laboratory, 
California Institute of Technology, under contract with the National  
Aeronautics and Space Administration.} to search for
known optical  counterparts to each of  the \HI\ sources  in the South
Celestial  Cap (SCC)  sample and  the HIPASS  Bright  Galaxy Catalogue
\citep[BGC;][]{kori04}, respectively.   On the other  hand, HOPCAT was
produced using an automated  visual interactive program which displays
\sex\ \citep{bertin96} ellipses  representing areas within the SUPERCOSMOS
fields which are above the sky intensity as well as velocities derived
from both 6dF and NED.   The automated visual display program was then
used  by three people to  interactively  compile HOPCAT.   In HOPCAT,  an
optical match  is proposed  when the HIPASS  velocity is  within 400
\kms\ of the velocity derived from NED/6dF and the positional match is
within the $15\arcmin \times 15\arcmin$ SUPERCOSMOS field.

The large Southern sky surveys used to generate HOPCAT, such as 6dF and
SUPERCOSMOS, are   not   available    for   the    construction   of
NOIRCAT. Northern analogues of 6dF and SUPERCOSMOS do not exist.
Most  large, recent  optical surveys  such as  SDSS do  not  cover the
entire Northern sky.  Hence, NED is used as our source catalogue for
optically matching the Northern HIPASS detections.  It should be noted
that we used the October 2006 version of NED.  To improve optical
detection limits, we also used the 2MASS near-infrared catalogue.
  For    NOIRCAT,    the   primary    method    for   determining    the
optical/near-infrared  matches  was  by  ``interactive''  cataloguing,
after  an automated  search  of  the NED  and  2MASS catalogues.   The
automated  search identified  all  the NED  and  2MASS sources  within
7.5\arcmin\ and 400 \kms\ of each NHICAT source.  These preliminary
search criteria  were intended  to be simple,  in order to  include
galaxies that may have extended \HI, or an \HI\ distribution that is 
offset from the optical counterpart.
These parameters are also consistent with HOPCAT's matching criteria.

The preliminary  matches were then plotted  with a `+'  and a number
onto  an optical  field  centred on  the  corresponding HIPASS  source
centre.  Figure~\ref{sample}  shows an example of  the graphics window
shown for HIPASSJ0150+21 (a Flag 3 source) during this interactive process.  
Each HIPASS field is  displayed  (with a list of properties 
found from the automated search),  inspected and  graded using  the
interactive process.    The  list  of  properties displayed include the  HIPASS
velocity,  velocity width,  name, optical  velocities (and  errors) as
well  as the  availability of  2MASS magnitudes  for each  of  the NED
sources found.  We obtained optical  fields for all the NHICAT sources
from the Second Palomar Sky Survey \citep[POSSII; ][]{reid91} in the red
band.  Table~\ref{flagtab} shows the five main categories into  which each source  was sorted. 

During the  interactive process,  the appropriate optical  matches and
match category are determined for  each HIPASS source.  The three rules
used for determining a match are:
\begin{enumerate}
\item Optical sources must be within 7.5\arcmin\ of the HIPASS centre. 
Where  there are multiple  source names referring to  the same
source (e.g.\ SDSS and APM nomenclatures), the non-SDSS/APM reference is
preferred.
\item  Optical velocity matches  are made  when the  published optical
velocity  (including velocity uncertainties)  is consistent  to within
100 \kms\ of the HIPASS velocity profile.
\item For optical velocities without  published errors in NED, a match
is recorded  when the published  optical velocity is  within
150 \kms\ of the HIPASS velocity profile.
\end{enumerate}

To be  consistent with previous HIPASS  optical counterpart catalogues
\citep{ryan02,doyle05}, we chose 7.5\arcmin\ to be the maximum angular
separation  between a  HIPASS source centre and  potential optical
match.   Although the predicted  position accuracy  is $\sim$3\arcmin,
the position accuracy also depends  on the \HI\ peak flux density, the
source extent and any asymmetries or confusion intrinsic to the source
\citep{barnes01}.    Previous  catalogues   have   also  found   match
separations greater  than 5\arcmin.  Within  HOPCAT, $0.6\%$ ($\sim25$
sources) of the HIPASS-optical  velocity matches correspond to angular
separations greater than 7.5\arcmin\ even though the average HIPASS beam
FWHM is 14.3\arcmin\ \citep{barnes01}.   Other reasons for
positional matches beyond $\sim$3\arcmin\ will be further discussed in
Section 3.


The  interactive   cataloguing  was  undertaken   by  two  independent
researchers (Wong and Waugh).  It should be noted that, unlike HOPCAT,
all multiple optical  matches (Flag 3 or 4) will  be listed in NOIRCAT
and no attempt has been made to choose between possible matches. In these cases,
discrimination can only be made with higher angular and velocity 
resolution \HI\ observations.

\subsection{Properties of NOIRCAT}

We  found  that 655  of  1002 NHICAT  sources  could  be matched  with
previously-catalogued  galaxies  for  which  an optical  velocity  was
available (Flags 1 to 4).  Of  these 655 sources, $82\%$ are matches to
single galaxies and  of these matches, $73\%$ have 2MASS observations 
in the  $J$, $H$ and $K$ wavebands.   Table~\ref{flagtab}   summarises  the  distribution  of
NHICAT sources over the seven match categories.  A full description of
NOIRCAT's  parameters and an example of the first 10 sources in NOIRCAT
can be found in Table~\ref{egnoircat}.  The total NIR $J$, $H$ and $K$ 
magnitudes from 2MASS are also listed in NOIRCAT.  Corrections\footnote{More
details on the 2MASS processing can be found at: {\tt{http://www.ipac.caltech.edu/2mass/releases/allsky/doc \newline /sec4\_5e.html$\#$large}}} were made to account for 
the 10\%--20\% flux loss due to the very high background levels in the 2MASS
observations.  

To examine the properties of  these matched galaxies, we use the 2MASS
$J$, $H$  and $K$ magnitudes instead  of the optical  magnitudes because the
2MASS catalogue is the best available optical/NIR catalogue and has
the best  Northern sky coverage  corresponding to the sky  coverage of
Northern HIPASS.  The  NIR wavelengths are also less  sensitive to the
dust obscuration  in the Galactic plane.  The  NIR apparent magnitudes
can  be  used as  stellar  mass  indicators.   The  NIR observations are
better tracers  of mass distribution  because NIR emission  is derived
primarily  from cooler  giant and  dwarf stars  (instead of  hot young
stars) which account for a major fraction of the bolometric luminosity
of a galaxy.   Using the NIR observations, we explore the relationships 
between  the \HI\ content  of HIPASS galaxies  and their inferred stellar 
content. 

\begin{figure}
\begin{center}
\includegraphics[scale=0.49]{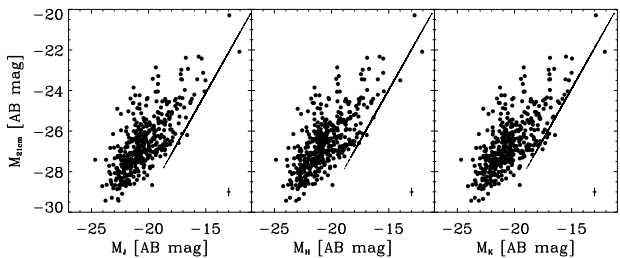}
\caption{HIPASS absolute magnitude (M$_{\rm{21cm}}$)  as  a function  of
$J$, $H$ and $K$ absolute magnitudes for the  414 Flag 1  sources.  The solid
 line in each plot shows the combined HIPASS limit for a detection that 
is five times the RMS and a 2MASS detection with a SNR of 10.  The average
uncertainty for our data is shown by the error bar at the bottom right corner 
of each plot.}
\label{sintjhk}
\end{center}
\end{figure}

Figure~\ref{sintjhk} shows the \HI\ absolute magnitude\footnote{The \HI\ 
absolute magnitude has been calculated using the AB magnitude system via:
$m=-2.5\;{\rm{log}}(f) - 48.6$ where $f$ is the measured \HI\ flux in units of 
ergs s$^{-1}$ cm$^{-2}$ Hz$^{-1}$} (M$_{\rm{21cm}}$) 
of Flag 1 sources as a function of the $J$, $H$ and $K$ absolute magnitudes.
The flux observation limit of Northern HIPASS is 0.07 Jy for a detection 
that is 5 times the RMS.  
 

{
\scriptsize{
\begin{landscape}
{\setlength{\textwidth}{8.5in}
\begin{table*}
\begin{center}
\caption{Example of  the first  10 sources of  NOIRCAT. The parameters are further
described below the table.   It  should be
noted that Flag 5b and 5c sources will not be listed in NOIRCAT.}
\label{egnoircat}
\begin{tabular*}{50pc}{lccccccccccc}
\hline \hline  
(1) HIPASS\_name & (2) RA\_HIPASS& (3) Dec\_HIPASS  & (4) Vel\_HIPASS &(5) W\_HIPASS & (6) Flag & (7) Optical\_source  & (8) RA\_optical \\  
&(9) Dec\_optical &(10) Separation &(11) NED\_type &(12) NED\_morph &(13) Vel\_optical &(14) Vel\_err\_optical& (15) Vel\_source  \\
 &(16) J &(17) J\_err  &(18) H &(19) H\_err  &(20) K &(21) K\_err  & \\
\hline
\hline  
HIPASSJ0001+05  &00:01:39.0 &  05:18:47  &  3956.9  &48.2 &  5&UGC12910 &00:01:28.4 \\
 & +05:23:22 &  5.3 & Galaxy &SB(r)m & 3949 &5&1990ApJS...72..245S \\ 
& --&--&--&--&--&-- & \\ 
HIPASSJ0002+16a &00:02:08.4 &  16:35:13 &1047.3  &108.5 & 5  & IC 5377  &00:02:05.4 \\
&+16:35:25  &0.7 &  Galaxy &  Im  & 1050  &4& 1990ApJS...72..245S\\  
&--&--&--&--&--&-- & \\ 
HIPASSJ0002+16b & 00:02:54.9 & 16:08:51 &1047.1 &456.8 & 1 & NGC  7814 & 00:03:14.9\\
&+16:08:44 & 4.8 & Galaxy& SA(S)ab & 1050 & 4 &  1999ApJS..121..287H \\
 & 8.089 & 0.020 & 7.361& 0.023  & 7.084  & 0.024 \\  
HIPASSJ0003+07 &00:03:46.1 &  07:29:02 &5240.5 & 181.9& 1 & NGC 7816 & 00:03:48.8\\
 & +07:28:43 & 0.8 & Galaxy& Sbc & 5241  & 5 & 1999ApJS..121..287H \\ 
& 10.561  & 0.024 & 9.863 &0.030 & 9.525 & 0.034 \\ 
HIPASSJ0003+15 & 00:03:58.4 & 15:11:59 &873.8&100.6 &  5 &UGC00017 & 00:03:43.3\\ 
 & +15:13:06& 3.8 &  Galaxy &Sm &878  &  6  &  1991RC3.9.C...0000d\\  
&  --&--&--&--&--&--  &  \\
HIPASSJ0004+07 &00:04:17.6 & 07:21:47 & 6193.1 &198.7 & 1 & NGC 7818 &00:04:08.8\\  
&   +07:22:46  &2.4  &  Galaxy   &  Scd  &   6201  &1  &1999ApJS..121..287H\\ 
& 12.117& 0.046& 11.499& 0.066 & 11.079& 0.069 &\\ 
HIPASSJ0004+20 &00:04:18.1 & 20:49:39  &2162.1 & 107.1& 1 &NGC 7817&00:03:58.9  \\  
&  +20:45:08  &  6.4  & Galaxy  &  SAbc  &2310  &1  &1999ApJS..121..287H\\ 
&  9.489& 0.006& 8.734 &0.007&  8.421& 0.008& \\
HIPASSJ0004+05 &00:04:34.4 & 05:50:00 & 3112.2 & 221.0& 1 & UGC00027 &00:04:28.8   \\   
&+05:50:50  &1.6&   Galaxy   &   Scd   &3118  &1   &1999ApJS..121..287H\\ 
& 13.275 &0.076  & 12.749 &0.120 & 12.265 &0.131&  \\  
HIPASSJ0006+17  &00:06:30.3  &  17:20:22 &875.4  &  78.4&  5  &UGC00047& 00:06:38.3\\ 
 & +17:17:03 &  3.8 & Galaxy  & Sdm &873 &  4 &1990ApJS...72..245S\\  
& --&--&--&--&--&--  &  \\ 
HIPASSJ0006+08&00:06:45.7  & 08:37:26  &5255.2&210.5  &1& UGC00052  & 00:06:49.5  \\
&+08:37:43 & 1.0 &  Galaxy &  SAc  &5257 &2  & 1999ApJS..121..287H\\  &
11.361& 0.037& 10.772& 0.053& 10.405& 0.060 & \\ 
\hline \hline
\end{tabular*}
\end{center}
NOTE.--- Parameter definitions. (1): HIPASS identification. (2): HIPASS right 
ascension (J2000). (3): HIPASS declination (J2000). (4): HIPASS heliocentric velocity 
at FWHM (\kms). (5): HIPASS velocity width (\kms). (6): Optical match category flag. 
(7): Source name of optical match. (8): Right ascension (J2000) of optical source.
(9): Declination (J2000) of optical source. (10): Spatial  separation  between the  HIPASS
source centre and the matched source ($\arcmin$). (11): NED's  classification  of  
source type  (e.g.\ Galaxy, Galaxy Pair). (12): Published NED morphology.
 (13): Velocity of optical source (\kms). (14): Velocity error  of optical  source (\kms).
(15): Reference for optical velocity. (16): $J$-band magnitude from total 2MASS flux.
(17): $J$-band magnitude error from 2MASS. (18): $H$-band magnitude from total 2MASS flux.
(19): $H$-band magnitude error from 2MASS. (20): $K$-band magnitude from total 2MASS flux.
(21): $K$-band magnitude error from 2MASS.
\end{table*}}
\end{landscape} }}

\begin{figure}
\begin{center}
\includegraphics[scale=0.69]{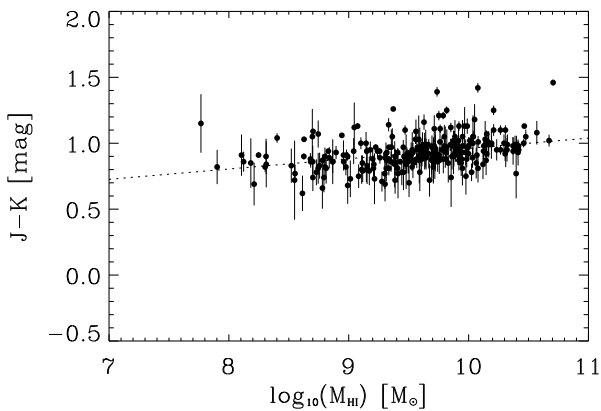}
\caption{$J-K$ colour as a function of  \HI\ mass for the Flag 1 sources with
reliable 2MASS colours.  The  dotted line shows the best robust 
linear fit to the data.}
\label{himass_colour}
\end{center}
\end{figure}

\begin{figure*}
\begin{center}
\includegraphics[scale=1.03]{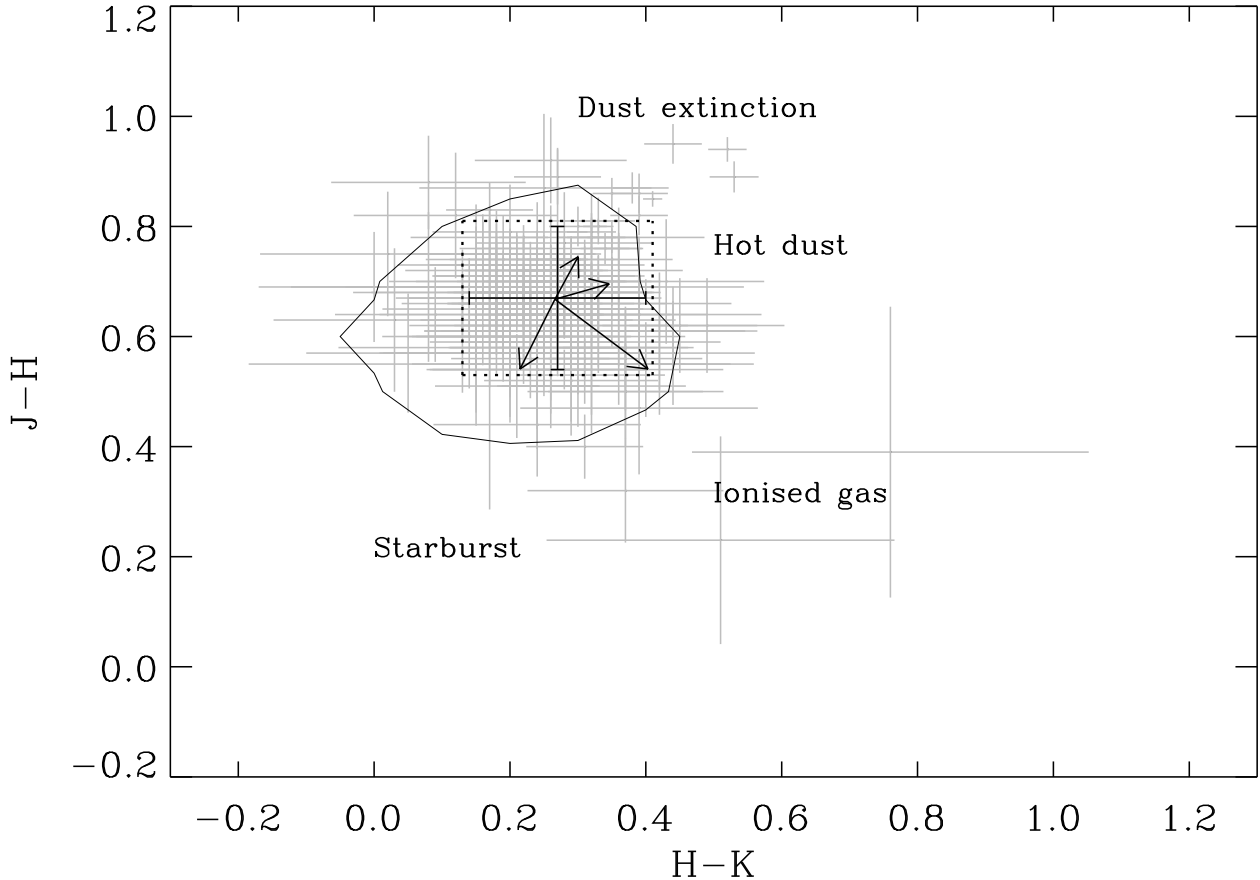}
\caption{$J-H$ versus $H-K$ colour-colour diagram for the  
Flag  1 sources with reliable 2MASS colours (in grey).  The black contours 
provide the 2-$\sigma$
contour  of the two-dimensional NIR colour distribution of our sample. The 
black cross marks the normal range of NIR colours (indicated by the error bars) 
for galaxies with nuclei dominated by an older stellar population. The black 
arrows indicate the shift in  direction of a galaxy's NIR colours due to factors 
such as starburst events,  gaseous emission from ionised regions, thermal 
re-radiation of hot dust and reddening \citep{geller06}.}
\label{colour_colour}
\end{center}
\end{figure*}

\noindent Likewise, the 2MASS observation limits are 
at log$_{10} (J) = -19.0$ W cm$^{-2}$ $\mu$m$^{-1}$, log$_{10} (H) = -19.2$ 
W cm$^{-2}$ $\mu$m$^{-1}$ and log$_{10} (K) = -19.4$ for a $SNR = 10$ 
detection in the $J$, $H$ and $K$ bands, respectively. A Pearson correlation
test resulted in an $R$-value of 0.72 for the relationship between all three
 NIR magnitudes with respect to M$_{\rm{21cm}}$.  This correlation at a 99.9\%
confidence level suggests that galaxies with more luminous in the NIR also 
have more \HI.   However, it should be noted that this correlation may be
 highly biased by the detection limits of the samples used and that at any
given \HI\ mass, there exists a broad range in NIR luminosity (as shown in
Figure~\ref{sintjhk}.

\citet{bell01} found a correlation between the stellar mass-to-light 
ratios and the colours of galaxies  from the 2MASS/SDSS passbands. 
 Using the NIR colours, we can distinguish between the galaxies 
which consist mainly of young stars, from the galaxies with larger fractions 
of older giants. The derived total NIR fluxes in 2MASS are too sensitive to 
stellar contamination and irregularities in the surface brightness
 profiles\footnote{{\tt{http://www.ipac.caltech.edu/2mass/releases/allsky/doc \newline /sec2\_3b.html}}} for the determination of NIR colours.  Therefore, 
to study the NIR colour properties of our sample, we use the NIR magnitudes
from the 2MASS isophotal photometry. More accurate NIR colours are
obtained, even though these flux values do not reflect the total flux of a source
since the isophotal measurements are set at the 20th magnitude per arcsecond squared
at the $K_s$ band (which roughly correspond to 1$\sigma$ of the background RMS).
 Figure~\ref{himass_colour} shows the $J-K$ colours (of 264 Flag 1 sources with
reliable 2MASS isophotal photometry in all three $J$, $H$ and $K$ bands) as
a  function of  the \HI\ mass ($M_{\rm{HI}}$).  Following \citet{roberts62}, the 
\HI\ mass was calculated from:
\begin{equation}
M_{\mathrm{HI}} = 2.356 \times 10^5 D^2 \int S\,dV \;,
\end{equation}
where  Hubble Flow distances (D) assuming $H_{\circ} = 73 $ \kms\ Mpc$^{-1}$
and \HI\ integrated fluxes ($\int S\,dv$) are used.  We found the best linear 
fit to be  $J-K  = 0.08\;  \rm{log}\, (M_{\rm{HI}})  + 0.18$.
Most of the observed scatter can be attributed to the uncertainties from the
2MASS automated processing pipeline \citep{jarrett07}. A good correlation was
found between the \HI\ mass and the NIR $J-K$ colours (Pearson $R$-value = 0.37), 
which suggests that galaxies with greater \HI\ masses appear to have redder $J-K$
colours at a 99.9\% confidence level.  This agrees with previous work 
\citep[e.g.\ ][]{hanish06} which found 
that galaxies with greater \HI\ mass correspond to galaxies with greater
optical $R$-band luminosity densities. Our results are also complementary to the
findings of \citet{bell03} that galaxies with redder optical colours 
have greater mass-to-light ratios.

Near-infrared studies of normal non-interacting galaxies with nuclei 
dominated by older stars found that such galaxies span a very narrow 
window in the $J-H$ versus $H-K$ colour-colour diagram 
\citep{geller06,giuricin93}.  Currently, the largest sample of NIR
colours of galaxy pairs is catalogued by \citet{geller06}.  The NIR 
properties of these interacting galaxies were compared to the NIR 
properties of normal galaxy population from the Nearby Field Galaxy 
Sample, NFGS (Jansen et al.\ 2000a,b).  A broader distribution of
 $J-H$ and $H-K$ colours were found for these interacting galaxies 
than for average field galaxies. \citet{geller06} interpreted this result 
as evidence for bursts of star formation (which shifts the NIR colours 
blueward) and for a dust-reddened/extincted and/or radiation from hot 
dust (which in turn results in redder colours).  Radiation from hot dust 
is thought to be responsible for the reddest $H-K$ colours.

Our \HI-selected sample of galaxies provides an interesting comparison
to these previous NIR studies based on optically-selected samples.
Figure~\ref{colour_colour} shows a NIR colour-colour plot of the same
sources as in Figure~\ref{himass_colour}.  The points from our dataset
are plotted in grey where 95\% of our sources lie within the  
colour distribution marked by the black solid contour.
 Also plotted on Figure~\ref{colour_colour} is 
a black cross indicating the  range of NIR colours for normal galaxies 
\citep{geller06,giuricin93}.  The region enclosed by the dotted-lines mark 
the range of NIR colours found from the NFGS used by \citet{geller06} as 
the benchmark of a normal field galaxy sample.  

A simple model was proposed by \citet{geller06} to explain qualitatively the
distribution of colours observed between the sample of galaxy pairs and the
NFGS sample.  Their model explained that:  (i) dust extinction will redden
the intrinsic colour of the galaxies, (ii) emission from bursts of star formation
will shift the NIR colours blueward, (iii) re-radiation from hot dust will in 
general redden the NIR colours, particularly the $H-K$ colours, and 
(iv) the emission from regions with ionised gas shifts the $J-H$ colours
blueward and the $H-K$ colours redward.    These effects are summed up 
in Figure~\ref{colour_colour} as vector arrows extending away from the 
median NIR colour of the normal field galaxies. 

As can be seen from Figure~\ref{colour_colour}, our dataset is not entirely
concentrated within the NIR colour region for the normal galaxies found by
\citet{geller06} and \citet{giuricin93}.   In total, 41 galaxies
 (15.5\%) of our sample have NIR colours outside the region bounded by the dotted lines in 
Figure~\ref{colour_colour}.  Of the outlying galaxies, 18 galaxies (6.8\%) have
$J-H < 0.54$ which suggests the existence of galaxies within our \HI-selected 
sample that exhibit the effects of star formation and gaseous ionising regions 
which appear proportionally under-sampled by the optically-based NFGS sample.  
Although bursts of star formation may be responsible for shifting the $H-K$ colours
bluewards, it is not clear why 6 galaxies (2.3\%) of our sample are found to 
have $H-K < 0.14$ and $J-H > 0.67$. No physical mechanism appear to be able to 
shift the $H-K$ colours blueward and the $J-H$ colours redward simultaneously. 

We attribute these unusual NIR colours to the
large uncertainties inherent in the measured isophotal magnitudes due to the 
high background levels of the 2MASS observations. Five of the
six galaxies have very low surface brightnesses (with $J$-band magnitudes
between 13.0 and 14.9), while the remaining source appears to have a very
variable background level. The 2MASS photometry pipeline is susceptible to 
deriving unphysical NIR colours in galaxies due to contamination by foreground
stars and variable background gradients$^5$.  

\begin{landscape}
{\setlength{\textwidth}{8.5in}
\begin{table*}
\caption{NHICAT properties  of 25 Flag 5b sources  found with probable
matches to sources not listed in NED.}
\label{twentythree}
\vspace{1pc}
\scriptsize{
\begin{center}
\begin{tabular}{lccccc}
\hline  
HIPASS\_name  & RA\_HIPASS  (J2000)  &  Dec\_HIPASS (J2000)  &Vel\_HIPASS (\kms\ )  & W\_HIPASS (\kms\ ) &Galactic Latitude ($^{\circ}$)\\  
\hline 
HIPASSJ0050+08 &00:50:06.4 & 08:37:51 & 9972.0 &124.1 & -54.24 \\ 
HIPASSJ0358+10 & 03:58:36.1 &10:03:19 & 1978.0  & 141.2 &-31.36\\ 
HIPASSJ0413+21 & 04:13:53.3  & 21:00:08 &3630.8 &150.3 &-21.33\\ 
HIPASSJ0426+18 & 04:26:41.1 & 18:29:16 & 3974.9 &64.7 &-20.72\\  
HIPASSJ0443+14   &  04:43:54.9  &  14:19:57  &   2726.3  &35.9  &-20.03\\
HIPASSJ0703+03   &  07:03:06.6   &   03:10:47  &   3548.8  &103.4  &4.04 \\
HIPASSJ0727+04   &  07:27:39.6   &   04:40:51  &   2087.0  &   120.9 & 10.18\\
HIPASSJ0758+10   &   07:58:12.7  &   10:59:36   &   2346.0  &   94.5&19.75\\
HIPASSJ0821+03b   &  08:21:45.9   &  03:21:04   &  4132.7   &252.8  &21.58\\
HIPASSJ0835+14&   08:35:15.8   &   14:15:07   &   5899.2   &176.7   &29.33\\
HIPASSJ0836+05&   08:36:37.5   &    05:14:54   &   1865.6   &60.4  &25.74 \\
HIPASSJ1025+20&   10:25:29.9   &    20:14:24   &   1209.4   &   59.6 &56.01\\
HIPASSJ1048+12a&10:48:00.8 &  12:13:37& 955.2 &136.8&57.49\\ 
HIPASSJ1154+12&11:54:16.0 &  12:26:29 & 1004.6 & 47.9& 70.12\\  
HIPASSJ1327+19& 13:27:14.3 &19:51:31 &  7152.4 &115.0 &79.04 \\  
HIPASSJ1515+05& 15:15:18.5 &  05:50:20 &677.7 & 93.1&49.73\\ 
HIPASSJ1551+08& 15:51:41.2 & 08:01:23 & 5121.2 & 52.1&43.32\\
HIPASSJ1917+11&   19:17:24.0   &    11:53:13   &   4542.4   &38.3  &-0.26 \\
HIPASSJ1919+14&   19:19:44.4   &   14:05:44   &   2811.9   &169.0 &0.27  \\
HIPASSJ1922+18&   19:22:50.1   &   18:42:50   &   3934.7   &   336.1& 1.79\\
HIPASSJ1937+09&   19:37:31.1   &    09:21:00   &   3148.2   &81.0  &61.23 \\
HIPASSJ1949+24&   19:49:58.4   &   24:10:50   &   3111.1   &253.6 &-1.05  \\
HIPASSJ2306+14&   23:06:07.3   &   14:43:10   &   1556.4   &109.0 &-40.97  \\
HIPASSJ2316+15& 23:16:54.1 & 15:47:44 & 4491.4 &78.5 &-41.33\\ 
HIPASSJ2347+06&23:47:25.0 & 06:47:59 & 3273.2 & 87.6& -52.70\\ 
\hline
\end{tabular}
\end{center}}
\end{table*}}
\end{landscape}

\begin{figure*}
\begin{center}
\includegraphics[scale=0.9]{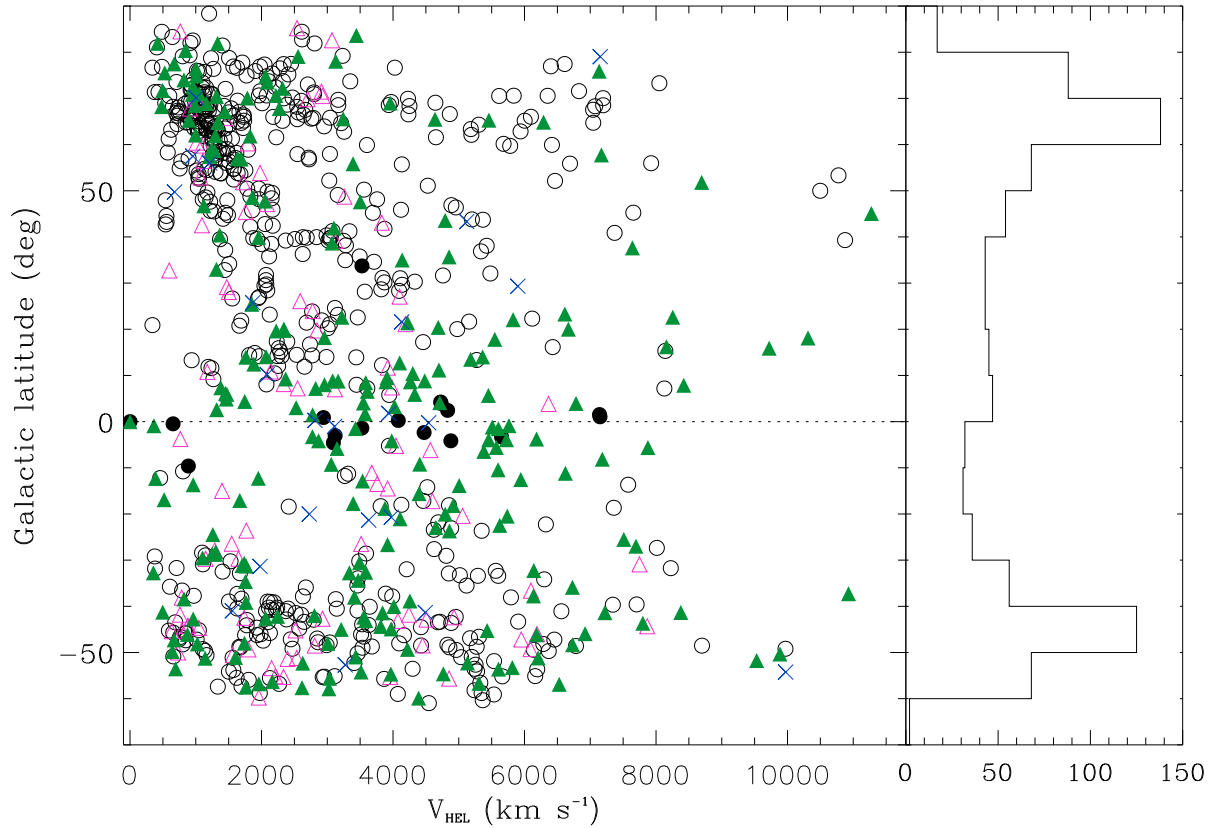}
\caption{The  galactic  latitude  of  each  match  as  a  function  of
heliocentric velocity  from NHICAT.  The black  open circles represent
the  NOIRCAT sources  in Flags  1,  2, 3  and 4.   The pink  open
triangles  represent  the  87 Flag 5a  sources  with  previous \HI\ velocity
matches.   The  green solid triangles and the blue crosses  represent  
the  219 Flag 5b sources (with probable matches based on positional 
matches) and the 25 category 5c sources (with probable matches with 
galaxies not listed  in NED), respectively.   The remaining  16 Flag 5c 
sources without any  matches to optically-visible galaxies  are represented by
the black solid  circles.  The cluster of sources  found at velocities
of  $\sim$1000   \kms\  and  at  Galactic   latitude  of  $65^{\circ}$
corresponds  to the  Virgo Cluster.  The cluster of sources 
at Galactic latitude of $-50^{\circ}$ is a result of projection effects
and is not the location of any known clusters. On the  {\em{right}}  panel, the
distribution of Galactic latitude of all the NOIRCAT sources is shown.
}
\label{gallatvel}
\end{center}
\end{figure*}

\noindent The masking of nearby foreground stars can be 
incomplete or not performed in some galaxies and therefore the resultant 
photometry for the galaxy will be corrupted.  Airglow gradients (from 
atmospheric OH airglow emission) in the background levels affect the accuracy 
of the measured photometry.  This airglow noise is known to vary strongly
 with time, spatial position and the size and total brightness of the extended 
source. In the $H$-band, the additional uncertainty introduced
by airglow is approximately equivalent to the measurement error\footnote{{\tt{http://www.ipac.caltech.edu/2mass/releases/allsky \newline/doc/sec1\_6c.html$\#$out}}}. 

 Therefore, the NIR colours of our \HI-selected sample of galaxies 
appear largely consistent with the NIR colours of the NFGS sample. 
 We attribute most of the observed scatter in NIR colours within our 
sample to the uncertainties in the 2MASS photometry.  Our sample may also
 include a greater fraction of star-forming galaxies exhibiting the effects of
 star formation and ionised gas than the NFGS sample.

\subsection{The search for dark galaxies}
\label{flag5}

For the 35\% of NOIRCAT sources without optical velocity matches (Flag 5), we
attempt to match these sources to sources from previous \HI\ surveys observed 
with a smaller spatial beamwidth than 7 arcminutes. For example, observations 
from the Very Large Array (VLA) with typical beamsizes below an arcminute will 
provide a suitable sample with which to compare. The combination of a more 
accurate pointing and the velocity of the \HI\ emission allow us to pinpoint the 
exact galaxy from which the emission is observed. Using the same interactive 
software algorithm detailed in Section 2.1, we  classify 87 optically-visible 
galaxies found with matching  \HI\ velocities as Flag 5a sources (see Table~\ref{flagtab}).



Of the remaining 260 Flag 5  sources, 219 have  one or more  positional 
matches with galaxies listed in NED (see Appendix~\ref{positonly}).  We 
identified these sources to be Flag 5b sources, and the rest of the 
sources (with no matches within 7.5\arcmin\ to galaxies in NED) were 
catalogued as Flag 5c sources.  We found 25 Flag 5c sources with possible 
positional matches to galaxies observed in the POSSII/DSS fields but not 
listed in NED (see Table~\ref{twentythree}).  The remaining 16 Flag 5c 
sources for which there were no optically-visible galaxies in  the POSSII 
fields are listed in Table~\ref{noopt}. One  of  the  sources (HIPASSJ0843+21) 
has  a bright  foreground star saturating  the field, while the other 15 
 are located in  crowded stellar fields in the direction of the Galactic  plane.
The mean Galactic latitude of these 15 sources is $-0.97^{\circ}$.


  A consistent result was found
by \citet{ryan02}, who found optical counterparts for their entire sample
except for one HIPASS BGC source, which was located
behind  the Large  Magellanic  Cloud  (LMC) where  the  field was  too
obscured  for any  identification.   Figure~\ref{gallatvel} shows  the
Galactic  latitude  of the  NHICAT  sources  as  a function  of  their
measured heliocentric velocity for different types of NOIRCAT matches.
Evidently, the  distribution of the  NOIRCAT galaxies is  dominated by
the substructure of the Local Universe.

In summary, no dark galaxy candidate has been identified in NOIRCAT. Since there
exists 244 galaxies for which a possible optical counterpart can be observed
in the POSSII fields, follow-up higher spatial resolution observations with velocity
information will pinpoint the exact location of the \HI\ source. Although we currently 
cannot rule out the existence of dark galaxy within our sample, the likelihood 
of their existence is small.

\section{Discussion}

\begin{figure}
\begin{center}
\includegraphics[scale=0.57]{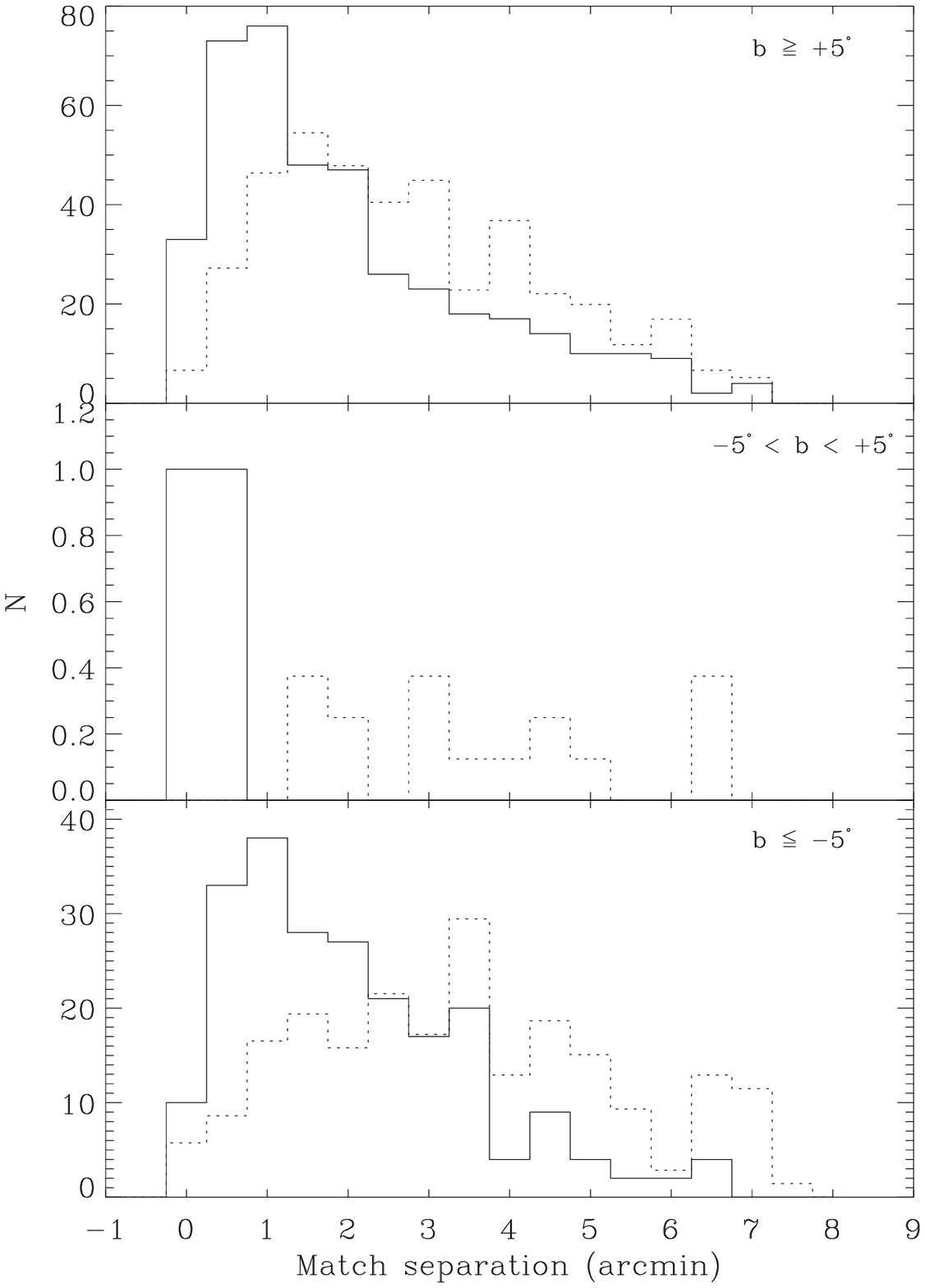}
\caption{Distributions  of  match  separations at three different Galactic 
latitude ($b$) ranges.  The top, middle and bottom panels show the match 
separations for sources located at $b \geq +5^{\circ}$, 
$-5^{\circ} < b < +5^{\circ}$ and $b \leq -5^{\circ}$, respectively. The
match separation distributions of the NOIRCAT matches (Flag 1, 2 \& 5a)
are represented by the solid histograms, while the dotted line histograms
show the distributions for the simulated sources which are one degree
away from any NHICAT source positions.  It should be 
noted that the NOIRCAT sample  had two sources in the range of 
$-5^{\circ} < b < +5^{\circ}$. In addition, the observed peaks in the
simulated distribution in the middle panel is due to the smaller 
sample size.}
\label{simsepdist_1deg}
\end{center}
\end{figure}

In this section, we examine the effectiveness of our identification 
process and the reliability of NOIRCAT. Can we statistically estimate 
how many of our optical matches are due to random matches?
We first determine the probability of identifying an optical source 
due to chance projection of positions on the sky by querying the 
NED database for 1,000 semi-random source positions chosen
one degree away in angular separation (in a random direction)
from any NHICAT source position in order to account for the structure 
and clustering of the Local Universe.  Figure~\ref{simsepdist_1deg} 
presents the sample-normalised distributions of the NOIRCAT match 
separations (solid line) and those of the semi-random matches (dotted 
line) divided into three Galactic latitude ranges. The match separation 
distribution for the semi-random matches have a larger average match 
separation than the distribution observed for the NOIRCAT matches. It 
should be noted that only the Flag 1, 2 and 5a NOIRCAT sources are 
represented in the solid line distributions shown in 
Figure~\ref{simsepdist_1deg}. The observed peaks in the middle panel of 
Figure~\ref{simsepdist_1deg} are due to the small sample size and are not 
real.

\begin{figure}
\begin{center}
\includegraphics[scale=0.57]{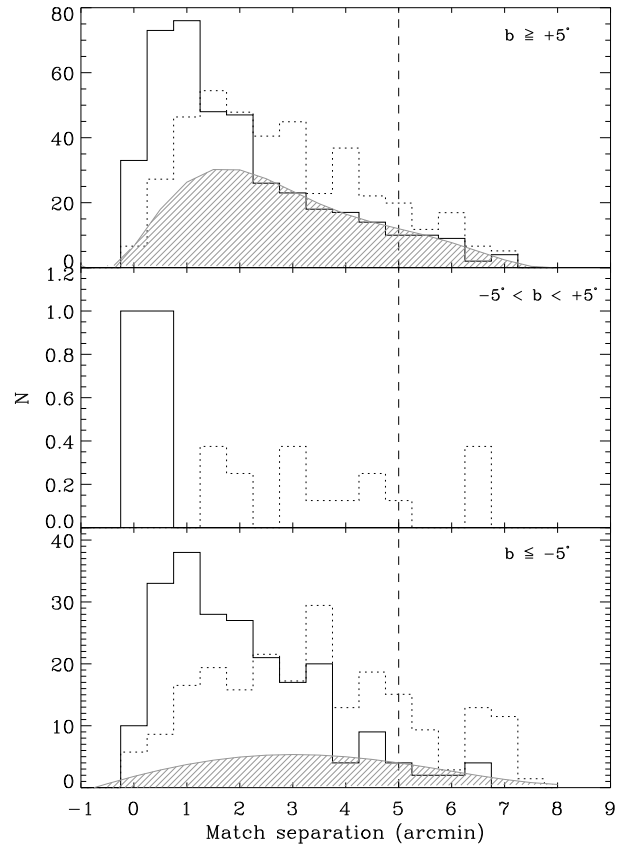}
\caption{Distributions  of  match  separations at three different Galactic 
latitude ($b$) ranges. See figure~\ref{simsepdist_1deg} for more details
on the distributions marked by the black solid and the black dotted lines.
The grey line-filled region shows the simulated random distribution scaled 
according to the tail end (at separations greater than five degrees) 
of the NOIRCAT match separation distribution. The vertical dashed line marks 
the match separation at five degrees in all three plots.}
\label{modeldist_1deg}
\end{center}
\end{figure}

Assuming that the real match separation distribution does not  have 
a tail and that all match separations greater than five arcminutes are random
matches, we fit the tail-end (where match separations is greater than 5
arcminutes) of the NOIRCAT match separation distributions
with a scaled-down distribution of the semi-random sample.  From this, we
can estimate a conservative upper limit on the possible number of random
matches within NOIRCAT if the NOIRCAT matching algorithm was entirely based 
on the two-dimensional positions of the objects on the sky. 
Figure~\ref{modeldist_1deg} shows this scaled model distribution of random
matches as a grey line-filled distribution which has been scaled-down from
the initial distribution of semi-random matches (shown in dotted lines). 
We find that 248 of the 408 matches in the Northern ($b \geq +5^{\circ}$) 
NOIRCAT distribution may be due to random matches. Similarly, 58 of 215 
matches in the Southern  ($b \leq -5^{\circ}$) NOIRCAT distribution may be 
due to random matches.  Each NOIRCAT match  along the Galactic equator
 ($-5^{\circ} < b < +5^{\circ}$) is real since the semi-random distribution is
completely different to the observed distribution.  Table~\ref{modeldisttab}
 provides a summary these results.  Therefore, the upper limit on the number
of NOIRCAT matches resulting from a chance projected alignment on the sky
({\em{if}} our matching algorithm was based solely on the positions of the 
sources) is 48\%.

\begin{table*}
\caption{NHICAT  properties  of the 16 Flag 5c sources without  optical
counterparts.}
\label{noopt}
\scriptsize{
\begin{center}
\begin{tabular}{lccccc}
\hline  
HIPASS name &  RA (J2000)  & Dec  (J2000)  & Vel$_{\rm{HEL}}$ (\kms\ )  & W  (\kms\ )  & Gal\_Lat (deg)  \\ 
\hline  
HIPASSJ0542+11 &05:42:43.6  & 11:27:29  & 887.3  &  109.4 &  -9.61\\ 
HIPASSJ0608+13  &06:08:35.7  & 13:06:50  & 5650.8  &  66.2 &  -3.30\\ 
HIPASSJ0636+04  &06:36:48.2    &     04:02:12    &    3526.1     &191.0    &-1.41    \\
HIPASSJ0843+21$^{\dag}$  &  08:43:14.6 &  21:29:23  &  3527.6 &  146.1&33.72\\ 
HIPASSJ1853+09  & 18:53:58.0  & 09:51:52 &  4731.7 &  331.9 &3.92\\ 
HIPASSJ1900+13 & 19:00:02.1 & 13:30:32 & 4724.8 & 96.7 & 4.25\\
HIPASSJ1901+06  & 19:01:35.4  &  06:52:00  & 2942.2  &  79.8 &  0.88\\
HIPASSJ1914+10  & 19:14:58.4  &  10:17:37  & 654.7  &  81.6 &  -0.47\\
HIPASSJ1919+18  & 19:19:53.7  & 18:47:37  &  4830.6 &  118.8 &  2.44\\
HIPASSJ1921+14  & 19:21:35.8  &  14:54:15  & 4080.3  &  81.7 &  0.25\\
HIPASSJ1922+08  & 19:22:10.4  & 08:13:21  & 3119.8  & 112.5  & -3.01\\
HIPASSJ1927+20  & 19:27:31.5  &  20:13:41  & 7141.0  &  72.4 &  1.53\\
HIPASSJ1929+08  & 19:29:09.1  & 08:06:27  & 3092.7  & 225.6  & -4.59\\
HIPASSJ1937+23  &  19:37:06.8 &  23:15:34  &  7148.5  &161.7 &  1.04\\
HIPASSJ1942+18  & 19:42:45.1  & 18:40:58  & 4473.0  & 127.0  & -2.36\\
HIPASSJ1950+18a  & 19:50:40.8  & 18:20:05  & 4879.6  &152.2  & -4.16\\
\hline
\end{tabular}
\end{center}}
\begin{flushleft}
$^{\dag}$  This source has  a foreground  star saturating  its optical
field.
\end{flushleft}
\end{table*}

\begin{table}
\caption{Conservative estimate on the number of random matches
which may have contributed to the final number of NOIRCAT matches {\em{if}}
the NOIRCAT matching algorithm was solely based on the two-dimensional
position of each object.}
\label{modeldisttab}
\begin{center}
\begin{tabular}{lccc}
\hline
Region & Total NOIRCAT matches & Real & Random\\
\hline
$b \geq +5^{\circ}$ & 408 & 162 & 246\\
$-5^{\circ} < b < +5^{\circ}$ & 2 & 2& 0\\
$b \leq -5^{\circ}$ & 215 & 161 & 54\\
\hline
\end{tabular}
\end{center}
\end{table}

\begin{figure}
\begin{center}
\includegraphics[scale=0.47]{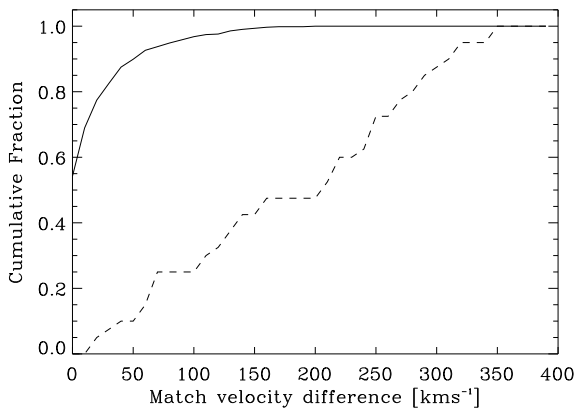}
\caption{Normalised cumulative distribution of match velocity offsets
for the NOIRCAT Flag 1, 2 and 5a sample (represented by the solid line) 
and the randomly-matched sample (represented by the dashed line).}
\label{velcompare}
\end{center}
\end{figure}

However, 74\% of NOIRCAT have matches in position as well as in velocity. The 
inclusion of velocity provides a better discriminant between real and 
spurious matches.  Although we use a very loose $\pm 400$ \kms\ velocity
offset to allow for possible matches, the average velocity difference (between
the HIPASS and optical/IR velocities) is 22 \kms.
To estimate the probability of a random velocity match, we query the NED 
database with 1,000 semi-random positions chosen one degree away in angular 
separation from a random NHICAT source position and set our random object's 
velocity at the velocity of the NHICAT source. Both position and velocity
will be used to determine a match. Even though a velocity offset of 
400 \kms\ appears very generous match criterion, we find that only 40 of 
our 1,000 semi-random sources were matched to objects within the NED
database. Figure~\ref{velcompare} compares the cumulative distributions
of the match velocity difference for NOIRCAT (solid line) and for the
randomly-matched sample (dashed line).  As shown 99.8\% of the NOIRCAT
sample with unique optical/IR source identification have match velocity
offsets below 200 \kms.  The average velocity offset of the semi-random
sample is 189 \kms.

For a conservative estimate of the number of random matches within
NOIRCAT, we now assume that the real velocity offset distribution does
not have a tail and that all velocity offsets greater than 100 \kms\ 
can be attributed to a random velocity match. Similar to the previous
test, we fit the tail-end (where velocity offsets are greater
than 100 \kms) of the NOIRCAT velocity offset distribution with a 
scaled-down distribution of the semi-random sample.  From this model
of semi-random matches, we find that there is a 1.5\% probability that
the Flag 1, 2 and 5a NOIRCAT sources may be due to random velocity matches.
In summary, if we only used the source positions as a matching criteria
then there is a 0.48 probability that a match is random. However, the 
inclusion of velocity as a match criteria significantly decreases the 
probability to 0.015 that a match made within NOIRCAT is spurious.

\begin{figure}
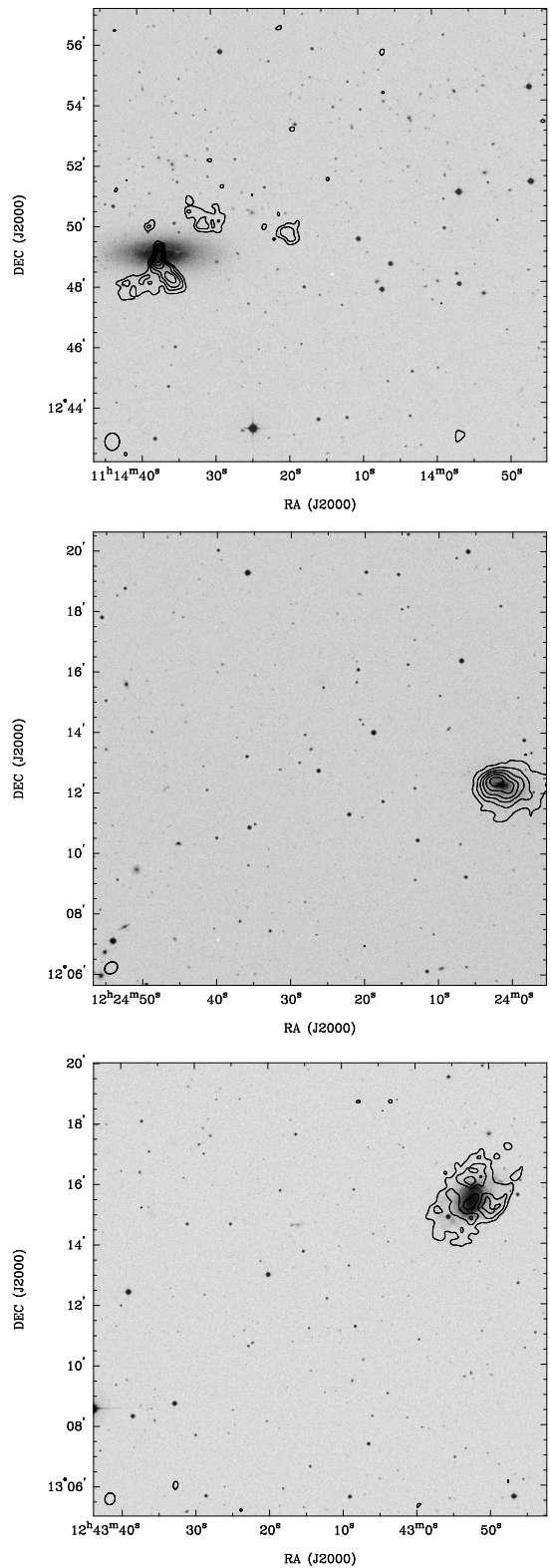

\begin{center}
\begin{tabular}{c}
\includegraphics[scale=0.35,angle=270]{fig9.ps}   \\
\includegraphics[scale=0.35,angle=270]{fig10.ps}\\
\includegraphics[scale=0.35,angle=270]{fig11.ps} \\
\end{tabular}
\caption{VLA integrated  flux contours overlaid on POSSII  images of 3
NOIRCAT  sources with  match  separations greater  than 5\arcmin.  The
centre  of the  fields correspond  to the  HIPASS  coordinate centres.
From  top to bottom: HIPASSJ1114+12  (NGC 3593),  HIPASSJ1224+12 (NGC
4351) and HIPASSJ1243+13a (NGC 4639).}
\label{vla}
\end{center}
\end{figure}
Although  the positional  accuracy  is 3\arcmin,  we found 36  Flag 1 and 2 sources
with match angular separations  greater than 5.0\arcmin.  Of these, we
found three with previous VLA observations listed in the NRAO
Archive\footnote{The  National   Radio  Astronomy  Observatory   is  a
facility of the National Science Foundation operated under cooperative
agreement  by Associated  Universities, Inc}.   Figure~\ref{vla} shows
the VLA  integrated flux contours  overlaid on POSS II  optical fields
for  these three NOIRCAT sources.   These three  sources (HIPASSJ1114+12,
HIPASSJ1224+12   \&  HIPASSJ1243+13a)   have   match  separations   of
5.3\arcmin, 6.1\arcmin\ and 6.3\arcmin, respectively. \citet{garcia00}
suggested  that HIPASSJ1114+12's  optical counterpart,  NGC  3593, had
accreted a gas-rich dwarf 1 Gyr  ago and recent stars had time to form
a central counter-rotating disk in the settling gas.  For both
HIPASSJ1224+12 and HIPASSJ1243+13a, the optical counterparts (NGC 4351
\& NGC  4639) are part of  the Virgo Cluster.  

Using the MBSPECT tool of the \miriad\ data reduction package, we measure
 the total integrated flux ($S_{\rm{INT}}$) and velocity width ($W_{50}$) 
from the VLA observations.  As detailed in Table~\ref{vlaprop}, the 
HIPASS integrated fluxes ($S_{\rm{INT}}^{\rm{HIPASS}}$) and the FWHM 
 velocity widths ($W_{50}^{\rm{HIPASS}}$)  are very different  to those 
measured from the VLA observations.

\begin{table}
\caption{Properties of VLA observations of HIPASSJ1114+12, HIPASSJ1224+12 
and HIPASSJ1243+13a.}
\label{vlaprop}
\scriptsize{
\begin{center}
\begin{tabular}{lcccc}
\hline 
HIPASS name &$S_{\rm{INT}}$ &$W_{50}$ & $S_{\rm{INT}}^{\rm{HIPASS}}$ & $W_{50}^{\rm{HIPASS}}$\\ 
 & Jy km s$^{-1}$& \kms & Jy km s$^{-1}$& \kms\\
\hline 
J1114+12 & 0.43&44.0& 14.1 & 215.1\\
J1224+12 &  8.79 &210.3& 5.3 &109.3\\
J1243+13a&1.93 &70.9& 35.2&331.0\\
\hline
\end{tabular}
\end{center}}
\end{table}

The \HI\ total integrated flux observed by the VLA of HIPASSJ1114+12 is 
less than $1\%$ of the total flux measured by HIPASS.  Additionally, the 
HIPASS velocity width is five times greater than the velocity width detected 
by the VLA. This suggests that most of the \HI\ emission measured by HIPASS 
in HIPASSJ1114+12 is diffuse and remains undetected by the VLA due to the 
lack of sensitivity at low column densities.  This can also explain the 
difference in the total integrated flux measured by HIPASS and the
VLA of HIPASS1243+13a.  

\begin{figure}
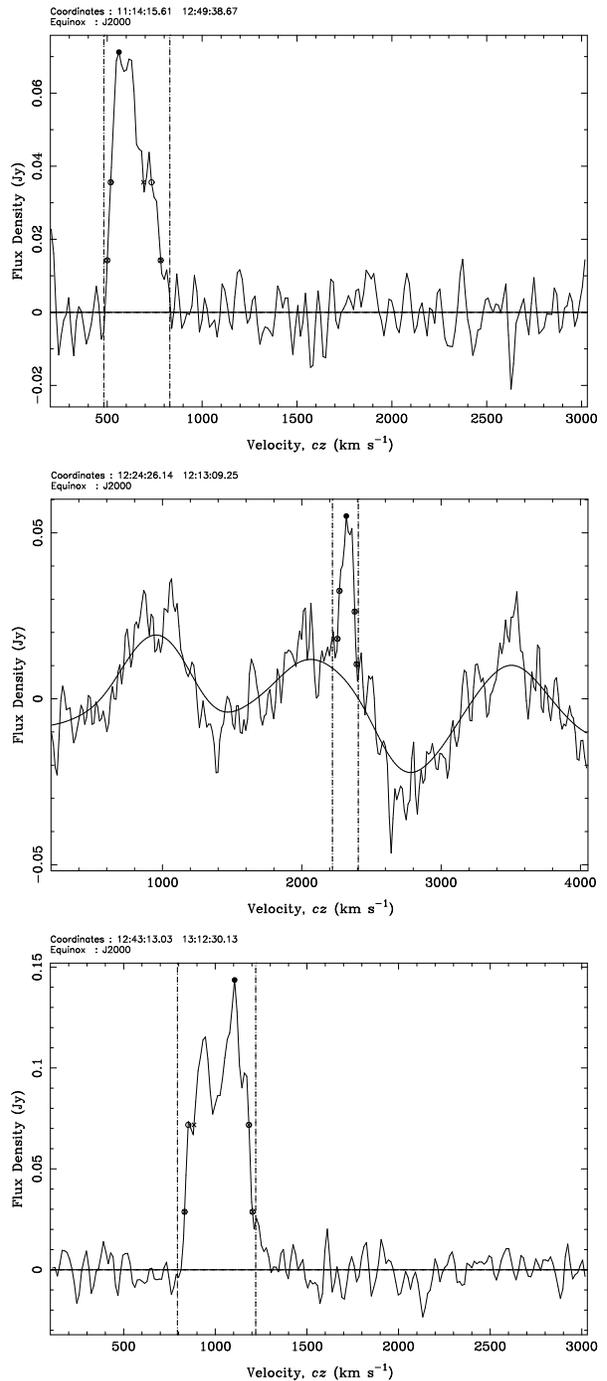

\begin{center}
\begin{tabular}{c}
\includegraphics[scale=0.32,angle=270]{fig12.ps}\\
\includegraphics[scale=0.32,angle=270]{fig13.ps}\\
\includegraphics[scale=0.32,angle=270]{fig14.ps}\\
\end{tabular}
\caption{HIPASS spectrum of HIPASSJ1114+12 (top), 
HIPASSJ1224+12 (middle) and HIPASSJ1243+13a (bottom).  The region 
between the dashed lines mark the \HI\ emission profile and the 
solid line shows the fit to the baseline.}
\label{spect}
\end{center}
\end{figure}

Conversely,  the VLA has observed more \HI\ integrated flux and a greater 
\HI\ velocity width than HIPASS in HIPASSJ1224+12. Due to the strong radio
emission from Virgo A (a Seyfert galaxy and a strong radio source), 
half of the emission profile was lost in the noise.    Thus, we would expect
that re-fitting the baseline would recover this flux. Figure~\ref{spect} 
shows the HIPASS spectra of the three sources listed in 
Table~\ref{vlaprop}.

\section{Summary}

In  this paper  we have  presented NOIRCAT,  the optical/near-infrared
counterpart catalogue  to NHICAT.  NOIRCAT  contains optically-matched
counterparts for  $65\%$ of the  NHICAT sources.  In  combination with
HOPCAT, NOIRCAT creates the largest catalogue of optical
counterparts of \HI\ sources, covering the entire sky in the declination range of
$-90^{\circ} < \delta < +25.5^{\circ}$.

Of the  347 Flag  5 sources, $24.5\%$  have optical  counterparts with
matching  velocities in previously-published  radio emission-line observations.
Another  $63.7\%$  have  probable  optical  counterparts  to  galaxies
without published velocities (other than HIPASS observations). 
Although our estimates in Section 3 showed that more than half these
matches should be real, only follow-up higher spatial resolution \HI\ 
observations of  these sources will help pinpoint the exact \HI\ position
 and constrain the possible number of dark  galaxies.  Many of 
the Flag 5c  sources lie in the direction of  the Galactic plane and as  such, 
are obscured behind our Galaxy.  

Our statistical analysis indicated that up to 1.5\% of the NOIRCAT
matches may be due to random matches. It is not possible to determine 
the exact number of matches from the current sample as more observations 
at better sensitivities and higher resolution are required.

Ignoring  the effects  of mergers,  \citet{verde02} postulated  that a
large  fraction   of  low-mass  halos   ($<  10^9$  \msun)   will  be
Toomre-stable and  not form  stars if the  gas collapse  during galaxy
formation  conserves angular  momentum.  In  addition,  simulations by
\citet{davies06} predicted  that `objects with scale sizes  of tens of
kpc  and  velocities of  a  few hundreds  of  \kms  can remain  dark'.
Contrary to  these results, \citet{taylor05} found that  a majority of
disks are  predicted to  be unstable  and likely to  form stars  in at
least half  the hypothetical dark galaxies with  baryon masses greater
than $5 \times 10^6$ \msun.  Standard reionisation models also propose
that dark halos do not contain gas \citep{susa04}.

\noindent Although our statistical analysis with our current dataset cannot confirm
the number of possible dark galaxy candidates within NOIRCAT, 
 detailed dark galaxy modelling by \citet{taylor07} on the NHICAT 
completeness limits found that there could be $\approx 10$ dark galaxies 
within NHICAT. These galaxies would be at the predicted limiting threshold
 for the formation of stars, if they formed. Hence, further observations 
of the Flag 5b and 5c sources may confirm the existence of these galaxies 
unless they are located in the direction of the Galactic plane.

As with  other HIPASS  catalogues, NOIRCAT will  be publicly-available
online  at: $\langle${\bf{\tt  http://hipass.aus-vo.org}}$\rangle$. 
We will also submit NOIRCAT to the NED database. Sources 
classified as Flags 1, 2, 3, 4 and 5a are  included  in the official NOIRCAT 
online.

\vspace{1cm}

\chapter{\bf{Acknowledgments.}}
We thank the anonymous referee for providing detailed and constructive 
comments which greatly improved this paper.
This  research has made  use of  the NASA/IPAC  Extragalactic Database
(NED) which  is operated by the Jet  Propulsion Laboratory, California
Institute of Technology, under  contract with the National Aeronautics
and Space  Administration.  This  research has also  made use  of data
products from the Two Micron All  Sky Survey, which is a joint project
of  the University of  Massachusetts and  the Infrared  Processing and
Analysis  Center/California  Institute of  Technology,  funded by  the
National Aeronautics and Space Administration and the National Science
Foundation. O.I.W. thanks  Marianne Doyle and Mike Read  for their help
in acquiring the POSSII image fields  as well as Stuart Wyithe for his
comments  and  ideas  about  dark  galaxies with  respect  to  current
reionisation models.

\nocite{*}
\bibliographystyle{mn2e} \bibliography{mn-jour,paperef}

\appendix

\section{NHICAT properties of 221 Flag 5b sources for which NED had 1 or more sources which were classified as galaxies with positional matches only}
\label{positonly}

\footnotesize{

\begin{table*}
\begin{center}
\caption{NHICAT properties of 219 Flag  5b sources for which NED had 1
or  more sources  which were  classified as  galaxies  with positional
matches only .}
\label{positonlytab}
\begin{tabular}{lcccc}
\hline \hline HIPASS\_name &  RA\_HIPASS (J2000) & Dec\_HIPASS (J2000)
& Vel\_HIPASS (\kms\ ) & W\_HIPASS (\kms\ ) \\ 
\hline 
HIPASSJ0003+08 & 00:03:14.6 & 08:42:34 &  2626.7 &139.1\\ 
HIPASSJ0003+15 & 00:03:58.4 &15:11:59 & 873.8  & 100.6 \\ 
HIPASSJ0010+13 &  00:10:42.8 & 13:43:57 &1740.5  &  76.4\\ 
HIPASSJ0014+07  &  00:14:39.8  &  07:30:58 &  3513.9&108.2\\  
HIPASSJ0016+07  &  00:16:54.8  & 07:12:06  &3967.5  &402.3\\
HIPASSJ0017+04   &   00:17:01.1   &   04:55:42   &   6528.2   &501.3\\
HIPASSJ0019+04   &   00:19:28.3   &   04:04:42   &   3023.8   &543.1\\
HIPASSJ0020+08 & 00:20:06.4 & 08:28:58 & 5604.5 &67.0\\ 
HIPASSJ0020+10& 00:20:03.0 & 10:53:16 & 1142.4&134.3\\ 
HIPASSJ0021+08 & 00:21:00.9 &08:35:06 &  693.1 &  41.3\\ 
HIPASSJ0028+11 &  00:28:54.3 &  11:18:15 &6207.9 &81.3\\ 
HIPASSJ0033+02 & 00:33:44.3 & 02:40:37 & 4389.2&225.9\\
HIPASSJ0109+13 & 01:09:57.9 & 13:18:37 & 4219.4&166.5\\ 
HIPASSJ0120+05&01:20:20.7  & 05:49:57 &  2165.0&82.8\\ 
HIPASSJ0121+12  &01:21:20.6 &12:25:35  & 642.7&119.1\\  
HIPASSJ0129+10  & 01:29:33.0  & 10:00:01  &9530.9 &93.3\\  
HIPASSJ0131+23 &01:31:21.7 &  23:54:22 & 3413.0&61.5\\
HIPASSJ0133+14 & 01:33:13.0 &  14:23:11 & 671.0 &78.0\\ 
HIPASSJ0134+04& 01:34:53.5 & 04:24:17& 1959.5 &108.1\\ 
HIPASSJ0142+02 & 01:42:28.4 &02:56:20  & 1763.9  &80.6\\  
HIPASSJ0143+19 &  01:43:15.9 &  19:58:20&496.7 &71.7\\  
HIPASSJ0158+04 &01:58:05.1 &  04:21:43 & 4765.3&312.4\\
HIPASSJ0210+06 &02:10:41.4 &  06:46:27 & 1604.0&103.3\\ 
HIPASSJ0211+14&02:11:49.9 &  14:11:42 & 3812.7&58.9\\ 
HIPASSJ0221+14  & 02:21:52.1 &14:19:27  &3586.3 & 51.7\\ 
 HIPASSJ0237+12 &  02:37:26.3 &  12:31:07 &960.2 &50.9\\ 
HIPASSJ0239+12 &  02:39:29.9 & 12:41:25 & 3554.1&154.2\\
HIPASSJ0243+16 & 02:43:16.8 &  16:45:37 & 821.4 &44.8\\ 
HIPASSJ0247+03&02:47:55.9 &  03:53:38 & 1024.6&90.5\\ 
HIPASSJ0250+03  & 02:50:56.6 &03:22:08  & 3009.1 &46.5\\  
HIPASSJ0251+06 &  02:51:32.0 &  06:02:30 &6921.3  &111.5\\  
HIPASSJ0253+02  &  02:53:48.6 &  02:20:42  &  6731.0&349.4\\  
HIPASSJ0253+06  &  02:53:09.3  & 06:32:03  &  5431.5&328.9\\
HIPASSJ0314+24   &   03:14:22.0   &   24:10:20   &   1303.8   &143.9\\
HIPASSJ0320+17 & 03:20:24.3  & 17:18:56& 355.2&74.2\\ 
HIPASSJ0332+15 &03:32:18.4 & 15:26:35 &  6141.0 &123.1\\ 
HIPASSJ0339+08 & 03:39:33.5 &08:31:37 &  6728.9 &110.3\\ 
HIPASSJ0340+05  & 03:40:51.5 &  05:22:08 &6137.2  &65.2\\  
HIPASSJ0341+24  &  03:41:19.2  &  24:00:54  &  1259.6&110.3\\  
HIPASSJ0341+18  & 03:41:53.5  &  18:07:21  & 1296.8  &45.3\\
HIPASSJ0345+08 & 03:45:11.7 & 08:51:12 & 1755.6 &40.9\\ 
HIPASSJ0345+02& 03:45:36.6 & 02:12:00 & 4256.7 &75.1\\ 
HIPASSJ0354+06 & 03:54:41.4 &06:37:15&  3470.4&249.8\\  
HIPASSJ0413+24 &  04:13:29.7  & 24:50:23  &3877.4  &130.3\\  
HIPASSJ0414+02   &04:14:24.5  &  02:46:54  &  3336.4&270.8\\  
HIPASSJ0415+02 &  04:15:41.3  & 02:28:45  & 3584.6  &140.6\\
HIPASSJ0417+13   &   04:17:55.1   &   13:30:32   &   7510.4   &396.3\\
HIPASSJ0421+10 & 04:21:10.1 & 10:09:26 & 7694.7 &93.6\\ 
HIPASSJ0428+18& 04:28:50.7 & 18:57:19 & 4793.7 &87.5\\ 
HIPASSJ0431+07 & 04:31:07.3 &07:24:28  &3917.2 &104.9\\  
HIPASSJ0432+16 &  04:32:42.8 &  16:12:18 &4107.1  &44.6\\  
HIPASSJ0446+08  &  04:46:31.4 &  08:18:58&  4651.7  &241.8\\ 
HIPASSJ0503+18  &05:03:14.4  &  18:24:56  &  5005.8  &389.1\\
HIPASSJ0506+25   &   05:06:50.7   &   25:12:46   &   3059.7   &254.5\\
HIPASSJ0508+10   &   05:08:49.5   &   10:45:00   &   1669.1   &555.1\\
HIPASSJ0519+22   &   05:19:44.1   &   22:56:38   &   7182.5   &314.9\\
HIPASSJ0520+17 & 05:20:50.9 & 17:02:24& 6621.8&168.8 \\ 
HIPASSJ0524+04&05:24:58.6 & 04:31:11 & 519.3 & 166.5\\ 
HIPASSJ0524+07 & 05:24:18.8 &07:23:46 &  4395.3 &106.7\\ 
HIPASSJ0527+15  & 05:27:42.9 &  15:52:52 &5598.2 &161.2\\ 
HIPASSJ0531+08 & 05:31:05.6 & 08:20:12 & 961.0 &90.1\\
\hline \hline
\end{tabular}       
\end{center}        
\end{table*}

\begin{table*}
\begin{center}
\contcaption{}
\begin{tabular}{lcccc}
\hline \hline HIPASS\_name &  RA\_HIPASS (J2000) & Dec\_HIPASS (J2000)
& Vel\_HIPASS (\kms\ ) & W\_HIPASS (\kms\ ) \\ 
\hline 
HIPASSJ0544+04 &05:44:24.9 &  04:13:03 &  3537.2 &66.7\\ 
HIPASSJ0545+05  &05:45:02.0 &05:04:09  & 387.8  &122.0\\  
HIPASSJ0547+17 &05:47:07.9  & 17:35:07  &5571.2&90.9\\
 HIPASSJ0554+18 & 05:54:10.1  & 18:00:42 & 5726.2 &46.7\\
HIPASSJ0556+13 & 05:56:26.8 & 13:40:07 &7877.4 &264.9\\ 
HIPASSJ0559+15& 05:59:53.0 & 15:36:00 & 5454.9 &196.8\\ 
HIPASSJ0603+08 &06:03:49.3 &08:38:43  & 5380.5  &206.6\\ 
HIPASSJ0605+19&  06:05:26.6 &  19:29:32 &5763.3  &286.2\\  
HIPASSJ0620+20&   06:20:36.1  &  20:10:49  &  1318.0&138.1\\   
HIPASSJ0621+11&06:21:18.5  &11:06:52   &   5602.4  &195.6\\
HIPASSJ0622+11&    06:22:49.2   &    11:08:28   &    5509.4   &384.1\\
HIPASSJ0623+04&    06:23:52.1   &    04:16:55   &    2867.8   &102.5\\
HIPASSJ0624+23&   06:24:35.3   &    23:21:20   &   1464.1   &   72.8\\
HIPASSJ0626+24& 06:26:39.8 & 24:40:18 & 1473.2 &99.2\\ 
HIPASSJ0630+23&06:30:04.1 &  23:34:08 & 1452.4 &135.4\\ 
 HIPASSJ0630+08& 06:30:09.2 &08:21:16  &  363.7 &53.4\\  
HIPASSJ0630+16&  06:30:08.5  & 16:47:50  &2526.4&275.1\\ 
HIPASSJ0631+02& 06:31:12.9 & 02:44:05 & 2774.1 &114.7\\
HIPASSJ0633+21&06:33:12.5 & 21:02:15 & 5451.5 &467.6\\ 
HIPASSJ0635+20&06:35:32.9 &  20:36:31 & 4329.3 &269.4\\  
HIPASSJ0635+11& 06:35:47.6 &11:13:11  & 3575.4  &76.6\\  
HIPASSJ0635+14 &06:35:52.7  & 14:36:39  &4023.3&347.5\\
 HIPASSJ0637+03& 06:37:39.3 & 03:24:50 & 3428.5 &159.8\\
HIPASSJ0637+22&06:37:56.1     &22:39:24      &     1380.5     &114.0\\
HIPASSJ0645+22&06:45:42.5     &22:25:58      &     4482.1     &195.0\\
HIPASSJ0656+06b&06:56:27.1 &  06:14:42& 6783.4&228.5\\ 
HIPASSJ0704+13&07:04:55.2 &  13:56:18 &  2367.5 &53.1\\ 
HIPASSJ0705+02&  07:05:43.6 &02:37:12  & 1744.6 &46.3\\  
HIPASSJ0710+05 &  07:10:09.9 &  05:16:33 &3609.1  &139.6\\  
HIPASSJ0714+06  &  07:14:00.6 &  06:18:00  &  8422.1&139.7\\ 
HIPASSJ0730+07 &  07:30:02.3 & 07:14:41 & 3920.0 & 94.0\\ 
HIPASSJ0731+08  & 07:31:16.0  &  08:00:01  & 1882.9  &55.2\\
HIPASSJ0755+03 & 07:55:52.3 & 03:27:05 & 9720.8 &93.4\\ 
HIPASSJ0818+04& 08:18:14.2 & 04:37:29  &4221.7 &52.2\\ 
HIPASSJ0831+07 & 08:31:31.0 &07:00:18 &  1850.8 &128.2\\ 
HIPASSJ0901+21  & 09:01:20.5 &  21:13:10 &7641.6  &86.5\\  
HIPASSJ0905+21  &  09:05:26.9  &  21:38:58  &  3073.6&107.0\\   
HIPASSJ0908+05a&09:08:12.9  &   05:55:02   &  1313.7&71.4\\
HIPASSJ0922+03&09:22:25.6 & 03:51:36 & 4139.3&128.9\\ 
HIPASSJ0942+04 &09:42:46.9 &  04:49:53 & 1955.0 &66.2\\ 
HIPASSJ1003+11  & 10:03:15.9 &11:29:33  & 3501.6 &52.8\\  
HIPASSJ1027+24 &  10:27:07.6 &  24:10:09 &1211.8 &52.5\\  
HIPASSJ1031+25 &10:31:34.9 &  25:16:07& 1282.3&125.4\\
HIPASSJ1034+23   &   10:34:38.9    &   23:03:06   &   1238.0   &49.3\\
HIPASSJ1052+07& 10:52:59.4  & 07:37:52 &  3392.0&80.2\\ 
HIPASSJ1106+19&11:06:02.6  & 19:49:15 &  1334.9&38.3\\ 
HIPASSJ1113+21  &11:13:43.0 &21:34:37  &  1440.2&164.6\\  
HIPASSJ1119+03  & 11:19:10.4  &  03:36:02&7169.7 &  202.7\\ 
HIPASSJ1119+09  & 11:19:16.4 &  09:34:56 &  995.2 &69.9\\  
HIPASSJ1122+13  &  11:22:33.2   &  13:40:32  &896.3  &  51.4\\
HIPASSJ1129+11 & 11:29:44.4  & 11:58:38 &3240.9 &94.6\\ 
HIPASSJ1130+23 & 11:30:48.2& 23:05:26  &  2912.0 &179.8\\
HIPASSJ1137+18& 11:37:29.7  & 18:22:18 & 946.0 &41.1\\  
HIPASSJ1148+23 &11:48:55.2 &23:48:54&  528.2&101.1\\  
HIPASSJ1204+16  &  12:04:00.7 &  16:30:43  &2063.9  &173.6\\ 
HIPASSJ1205+21  & 12:05:10.1  & 21:29:07  &  3128.5 &37.3\\   
HIPASSJ1209+14&12:09:56.3   &   14:20:57  &   820.0   &52.6\\
HIPASSJ1213+16 & 12:13:03.1 & 16:11:21 &7135.9 & 83.5\\ 
HIPASSJ1214+09&12:14:41.5 &  09:11:45 & 1784.3&121.9\\  
HIPASSJ1215+09b&12:15:53.0 &09:40:01&  2219.4  &35.6\\  
HIPASSJ1215+12  & 12:15:26.5  &  12:59:54&2090.3&36.0\\  
HIPASSJ1218+07&12:18:23.0 &  07:39:09  & 3956.6&150.4\\
HIPASSJ1219+06b&12:19:53.6 & 06:39:42  & 480.7&41.5\\ 
\hline \hline
\end{tabular}       
\end{center}        
\end{table*}

\begin{table*}
\begin{center}
\contcaption{}
\begin{tabular}{lcccc}
\hline \hline HIPASS\_name &  RA\_HIPASS (J2000) & Dec\_HIPASS (J2000)
& Vel\_HIPASS (\kms\ ) & W\_HIPASS (\kms\ ) \\ 
\hline 
HIPASSJ1230+09 &12:30:22.3  &  09:31:28 &  495.9  &69.5\\ 
HIPASSJ1230+17&12:30:06.6  &17:24:52 & 2556.1&282.9\\ 
HIPASSJ1231+20 &12:31:44.5  & 20:23:28  & 1332.1  &68.6\\ 
HIPASSJ1234+15&12:34:41.7 &15:12:11  &   671.1&91.2\\  
HIPASSJ1240+13&  12:40:01.5   &  13:50:48&1004.3&40.2\\  
HIPASSJ1242+05&12:42:43.3   &05:46:12  &  989.1&106.0\\
HIPASSJ1243+07&     12:43:13.1    &    07:37:56     &    1314.7&66.8\\
HIPASSJ1244+12&12:44:07.0 & 12:08:52 & 1008.1&111.7\\ 
HIPASSJ1250+17 &12:50:22.7 &  17:30:18 & 843.5 &116.9\\ 
HIPASSJ1256+19  & 12:56:04.9 &19:07:38  & 418.4  & 33.0\\  
HIPASSJ1314+23 &13:14:11.2  &  23:11:30 &3441.9&76.6\\  
HIPASSJ1327+10&13:27:18.7  &  10:03:17 &  1050.0&50.8\\
HIPASSJ1336+08   &   13:36:04.2   &   08:51:39   &1159.4   &   130.8\\
HIPASSJ1355+17&13:55:22.9 & 17:47:24  & 956.7 &97.5\\ 
HIPASSJ1403+09 &14:03:21.3 & 09:25:34 &  4638.5 & 43.9\\ 
HIPASSJ1404+08a& 14:04:15.4 &08:47:43  &  6289.2  &55.2\\  
HIPASSJ1406+22&14:06:56.0 &  22:04:40  &2320.0&122.3\\  
HIPASSJ1415+16&14:15:38.2 &  16:32:48  & 2270.6&90.0\\
HIPASSJ1420+08 & 14:20:54.4 & 08:40:24& 1298.0 &81.9\\ 
HIPASSJ1435+05&14:35:24.0 &  05:17:26 &  1636.3&96.0\\ 
HIPASSJ1435+13 &  14:35:37.9 &13:02:19  & 1827.0  &227.9\\ 
HIPASSJ1436+21  & 14:36:36.3  & 21:02:51&5458.5 & 291.9\\ 
HIPASSJ1445+07&14:45:16.0 & 07:52:49 & 1690.6& 38.3\\
HIPASSJ1526+14   &   15:26:22.1   &   14:25:11&   8694.9   &   136.8\\
HIPASSJ1538+12a&15:38:18.2       &       12:58:47&      1860.5&158.2\\
HIPASSJ1545+12&15:45:42.6 & 12:30:40 & 1122.5&110.4\\ 
HIPASSJ1548+16 &15:48:58.1  &  16:43:14&  2051.9 &94.5\\  
HIPASSJ1557+14&15:57:56.1  &14:58:41  & 11276.2&62.4\\  
HIPASSJ1604+14 &  16:04:10.8 &  14:37:54 &4793.0  & 111.5\\  
HIPASSJ1606+08 &  16:06:14.9 &  08:29:46&  1362.9 &145.9\\  
HIPASSJ1614+02  &   16:14:14.1  &  02:30:46&  4852.8  &78.3\\
HIPASSJ1621+20   &   16:21:38.7   &   20:52:25&   3100.6   &   128.1\\
HIPASSJ1735+02   &   17:35:34.7   &   02:46:46  &   10311.9   &139.3\\
HIPASSJ1736+15 & 17:36:49.7  & 15:12:00& 6613.4 &45.2\\ 
HIPASSJ1747+04& 17:47:06.5 & 04:11:58 & 8158.7 &80.9\\ 
HIPASSJ1747+18 & 17:47:31.8 &18:18:00  & 5827.3 &57.7\\  
HIPASSJ1750+21 &  17:50:13.4 &  21:15:50 &3221.7  &82.4\\  
HIPASSJ1752+22  &  17:52:56.0  &  22:54:44  &  8255.3&143.2\\  
HIPASSJ1754+02  &17:54:40.9   &  02:55:07  &  1761.7&187.1\\
HIPASSJ1758+14   &  17:58:46.7   &   14:47:43  &   2957.5  &   133.8\\
HIPASSJ1804+21    &18:04:37.7   &    21:39:04   &    2224.0&   196.6\\
HIPASSJ1805+17&18:05:05.9     &      17:19:09     &     5548.3&149.7\\
HIPASSJ1805+23a&18:05:31.2 & 23:13:00 & 2339.6&101.4\\ 
HIPASSJ1805+23b&18:05:57.2 & 23:26:38 &  6666.8 &58.6\\
HIPASSJ1807+09 & 18:07:57.2 &09:46:19&  2070.0&150.5\\  
HIPASSJ1807+25 &  18:07:22.0  & 25:19:41  &4688.0  & 126.2\\  
HIPASSJ1817+14  & 18:17:07.7  &  14:25:02 &  5362.7&200.5\\  
HIPASSJ1819+14 &  18:19:56.8  & 14:40:45  & 5181.1  &132.5\\
HIPASSJ1828+06   &   18:28:49.8   &   06:32:26   &   2957.3   &158.5\\
HIPASSJ1832+06 & 18:32:12.6 & 06:25:06 & 2823.3 &90.6\\ 
HIPASSJ1833+10& 18:33:29.4 & 10:38:48  & 3165.1 &158.1\\ 
HIPASSJ1837+11 & 18:37:37.3& 11:55:17  & 3584.4 &80.9\\  
HIPASSJ1837+22 &18:37:45.3 &  22:04:32 &4101.0&132.2\\  
HIPASSJ1839+13  &   18:39:35.3  &  13:19:21  &  3892.9&151.4\\  
HIPASSJ1842+17  & 18:42:53.8  &  17:02:14  & 3909.8  &89.4\\
HIPASSJ1842+15   &   18:42:58.3  &   15:00:29   &   4263.6  &   36.6\\
HIPASSJ1843+19   &   18:43:27.2   &   19:26:41   &   4300.7   &206.1\\
HIPASSJ1846+22    &18:46:11.3     &    22:36:24    &    4698.6&316.5\\
HIPASSJ1849+18&18:49:55.0 &18:44:16  & 3081.8&130.0\\ 
HIPASSJ1912+13 &19:12:37.3 & 13:23:54 &  2776.7 &112.2\\ 
HIPASSJ1915+20 & 19:15:00.4 &20:11:43 &  4710.3 &545.6\\ 
HIPASSJ1917+04  & 19:17:39.1 &  04:27:21 &6182.8 &167.2\\  
\hline \hline
\end{tabular}       
\end{center}        
\end{table*}

\begin{table*}
\begin{center}
\contcaption{}
\begin{tabular}{lcccc}
\hline \hline HIPASS\_name &  RA\_HIPASS (J2000) & Dec\_HIPASS (J2000)
&   Vel\_HIPASS   (\kms\   )   &   W\_HIPASS  (\kms\   )   \\   
\hline
HIPASSJ1950+18b&19:50:52.8 & 18:23:51& 3979.2&310.9\\
HIPASSJ2004+07&20:04:12.1 & 07:23:37 & 5943.3 &145.0\\ 
HIPASSJ2004+14&20:04:46.0 & 14:06:19 & 4406.0&138.4\\ 
HIPASSJ2015+12 &20:15:50.3 &  12:40:44 & 1950.9 &51.6\\ 
HIPASSJ2042+07  & 20:42:16.7 &07:36:53 &  5737.4 &120.5\\ 
HIPASSJ2045+12  & 20:45:19.4 &  12:53:03 &4916.0  &124.2\\  
HIPASSJ2109+21  &  21:09:41.2 &  21:18:16  &  3394.7&111.2\\   
HIPASSJ2112+12&21:12:26.7  &   12:37:20   &  4857.6&200.0\\
HIPASSJ2132+07&21:32:52.3 & 07:58:22  & 3490.8&66.7\\ 
HIPASSJ2142+22 &21:42:27.7  &  22:38:58  &5623.0 &84.4\\  
HIPASSJ2149+14&21:49:34.4  &14:14:35   &  1103.0&119.4\\  
HIPASSJ2158+14&21:58:34.9   &  14:07:01&1703.2&90.5\\  
HIPASSJ2207+15&22:07:06.0 &  15:59:05 &  1767.1 &64.3\\
HIPASSJ2208+03   &   22:08:05.5   &   03:36:27   &   4012.2   &152.7\\
HIPASSJ2209+01   &   22:09:44.7   &   01:59:07   &   3840.3   &100.7\\
HIPASSJ2211+17&22:11:54.7  &17:54:35 &  1738.0&151.3\\ 
HIPASSJ2224+22&22:24:52.1 &  22:58:23 &  1249.8&90.6\\ 
HIPASSJ2225+06 &  22:25:31.1 &06:23:15  &  8377.6  &69.1\\  
HIPASSJ2251+07&22:51:29.7 &  07:15:58  &3211.6&105.6\\  
HIPASSJ2253+11  & 22:53:48.0  &  11:16:15  & 2243.5  &148.8\\  
HIPASSJ2255+11  & 22:55:37.8  &  11:03:54  & 2064.3  &146.8\\
HIPASSJ2301+12& 23:01:06.0 &  12:44:53 & 2805.2&213.5\\ 
HIPASSJ2308+17& 23:08:47.5 & 17:12:44  & 1764.2 &142.0\\ 
HIPASSJ2316+05&23:16:00.0 &05:11:47  &  9884.7&152.9\\  
HIPASSJ2319+16&23:19:36.8  &  16:06:43  &7222.5&478.0\\ 
HIPASSJ2322+13 &  23:22:27.2 & 13:52:11 &7799.7 &64.1\\
HIPASSJ2333+04   &   23:33:20.8    &   04:23:33   &   5814.1   &88.9\\
HIPASSJ2336+12&23:36:29.9      &     12:46:22      &     6184.4&62.1\\
HIPASSJ2336+14&23:36:41.6 & 14:12:05 & 3971.0&174.2\\ 
HIPASSJ2339+07 &23:39:32.7  & 07:48:53  & 3430.9  &60.5\\  
HIPASSJ2349+02&23:49:53.6 &02:43:12  &  5307.8&222.8\\  
HIPASSJ2353+07&23:53:55.1  &  07:56:25  &5132.9&167.3\\  
HIPASSJ2357+23  &  23:57:57.2  &  23:59:54  &  10927.8&100.2\\  
HIPASSJ2358+04 &  23:58:15.2  & 04:47:37  & 3035.2  &110.7\\
HIPASSJ2359+02 & 23:59:17.3 & 02:42:05 & 2616.0 &147.8\\ 
\hline \hline
\end{tabular}       
\end{center}        
\end{table*}        }

\bsp

\label{lastpage}

\end{document}